\documentclass{ifacconf}
 
\usepackage{natbib}        

\usepackage{tikz}
\usetikzlibrary{calc}
\usetikzlibrary{math}
\usetikzlibrary{shapes,arrows,calc,angles,quotes,arrows.meta}
\usepackage{pgfplots}
\usepgfplotslibrary{fillbetween}
\pgfplotsset{compat=1.17}
\usepackage{soul}
\usepackage{amssymb}
\usepackage[cmex10]{amsmath}
\usepackage{graphicx}
\usepackage{cases}
\usepackage{enumerate}
\usepackage{epstopdf}

\usepackage{array}
\usepackage{multirow}
\usepackage{bigdelim}
\usepackage{mdwtab}
\usepackage{physics}
\usepackage{color}

\usepackage{bm}
\newcommand{\bx}{\bm{\mathrm{x}}}
\newcommand{\blambda}{\bm{\mathrm{\lambda}}}
\DeclareMathOperator{\sign}{sgn}

\newcommand{\x}[3][]{%
  \ensuremath{\mathcal{X}_{#3}^{#1}}%
}

\usepackage{accents}

\newtheorem{remark}{Remark}[section]
\newtheorem{assumption}{Assumption}

\begin{document}
\begin{frontmatter}
\title{Minimum-Time Escape from a Circular Region for a Dubins Car} 
\author{Timothy L.~Molloy}
\author{and Iman Shames}
\address{CIICADA Lab, School of Engineering, The Australian National University (ANU), Canberra, ACT 2601, Australia\\
  (e-mail: \{timothy.molloy,iman.shames\}@anu.edu.au)}

\begin{abstract}
We investigate the problem of finding paths that enable a robot modeled as a Dubins car (i.e., a constant-speed finite-turn-rate unicycle) to escape from a circular region of space in minimum time.
This minimum-time escape problem arises in marine, aerial, and ground robotics in situations where a safety region has been violated and must be exited before a potential negative consequence occurs (e.g., a collision).
Using the tools of nonlinear optimal control theory, we show that a surprisingly simple closed-form feedback control law solves this minimum-time escape problem, and that the minimum-time paths have an elegant geometric interpretation.\\[0.2cm] © 2022 the authors. This work has been accepted to IFAC for publication under a Creative Commons Licence CC-BY-NC-ND
\end{abstract}

\end{frontmatter}

\section{Introduction}
\label{sec:introduction}
The possibility of ubiquitous mobile robots and autonomous vehicles in the air, in the sea, and on land, has led to renewed interest in a variety of path planning problems and solution techniques \citep{Mac2016,Lavalle2006,Burke2020,Matveev2020,Molloy2020,Shima2020, Chitsaz2007}.
Specific path planning problems that have attracted recent attention include collision avoidance (cf.\ \citep{Molloy2020,Johansen2016} and references therein), interception \citep{Buzikov2022,Molloy2020}, and minimum-distance or minimum-time travel \citep{Boissonnat1994,Bui1994,Shima2020,Chen2020,Chitsaz2007}.
Despite the variety of path planning problems previously considered, the fundamental problem of how to turn a mobile robot with a finite maximum turn rate and constant speed so that it leaves a fixed circular region of space in minimum time appears to have not previously been posed or solved.
This minimum-time escape problem arises in situations such as collision avoidance \citep{Molloy2020} and capture-the-flag \citep{spenceropposed} in which a robot may enter regions of space to perform a task but must then exit them quickly.
More intricate path planning problems are also often decomposable into sub-problems that resemble this minimum-time escape problem (e.g.,\ the Dubins traveling salesman problem \citep{Isaiah2015}).
In this paper, we therefore seek to pose and solve this minimum-time escape problem using analytic techniques from nonlinear optimal control theory.

Path planning for vehicles that have a finite maximum turn rate is challenging because their physically realizable paths are subject to curvature constraints.
Planning minimum-time paths between two points under curvature constraints has a long history, going back at least to \cite{Markov1887}, and attracting widespread attention following the work of \cite{Dubins1957,Dubins1961}.
In particular, \cite{Dubins1957,Dubins1961} introduced the \emph{Dubins car} model of a vehicle with constant speed and finite maximum turn rate, and showed that minimum-time paths between two specific points for such vehicles only consist of combinations of straight lines and maximum-rate turns left and right.
Dubins' results have since been further developed using geometry \citep{ayala_geometric_2018} and nonlinear optimal control theory, specifically Pontryagin's principle \citep{Boissonnat1994,Bui1994,Chitsaz2007,Shima2020,Manyam2019,Chen2020}.
However, even with the tools of nonlinear optimal control theory, it has proved surprisingly difficult to analytically characterize minimum-time solutions for Dubins cars in terms of \emph{optimal feedback control laws} (i.e., expressions for the optimal controls as functions of system state).
Most existing solutions are thus described only as solutions to polynomial equations (cf.\ \citep{Shima2020}).

Most recently, the problem of finding a minimum-time path for a Dubins car from a fixed point to any point on a circle with the final heading tangent to the circle has been investigated in \citep{Manyam2019,Chen2020}.
Whilst solutions to this problem with elegant analytic descriptions have been identified in \citep{Chen2020} using Pontryagin's principle, they currently lack expressions in terms of feedback control laws.
The solutions of \citep{Chen2020} also provide limited insight into the closely related problem we consider in this paper of finding a minimum-time path from a fixed point \emph{inside} a circle to any point on the circle \emph{without} requiring that the final heading is tangent to the circle.

The key contribution of this paper is the full analytic solution of the problem of finding a path that enables a Dubins car to exit a circular region of space in minimum time.
By developing a bespoke argument built upon Pontryagin's principle, we identify a simple time-optimal closed-form feedback control law.
Besides the recent work of \cite{Chen2020}, the closest work appears to be the two-player surveillance-evasion game of \cite{Lewin1975} in which one player seeks to escape a circle whilst the other attempts to prevent escape.
However, in this surveillance-evasion game, the player tasked with escaping has an infinite maximum turn rate (i.e., can change heading instantaneously), and thus is much less constrained in its motion than the (nonholonomic) Dubins car we consider in this paper.
Furthermore, the solutions identified for the surveillance-evasion game of \cite{Lewin1975} lack elegant analytic or feedback descriptions.

This paper is organized as follows.
In Section \ref{sec:problem}, we formulate our minimum-time escape problem.
In Section \ref{sec:ocsol}, we present an approach to solving our escape problem using Pontryagin's principle, and leading to a feedback control law.
In Section \ref{sec:geo}, we provide deeper geometric insight into our solution.
Finally, in Section \ref{sec:simulations}, we present simulations illustrating minimum-time escape paths before discussing conclusions and future work in Section \ref{sec:conclusion}.

\section{Problem Formulation}
\label{sec:problem}

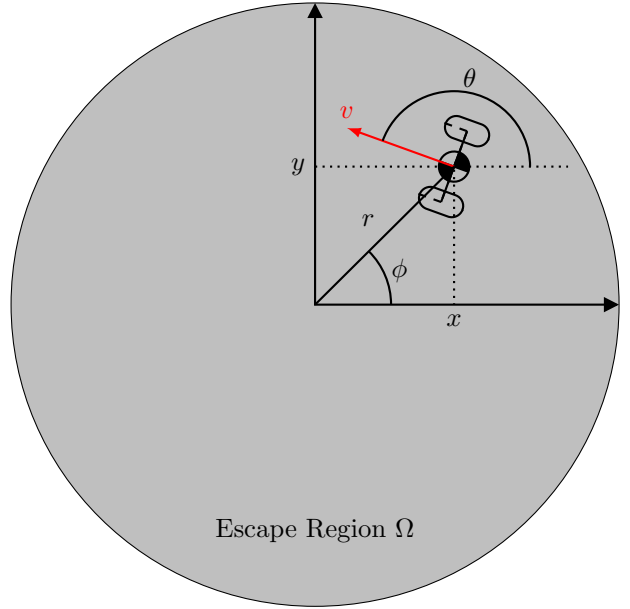
\begin{figure}[t!]
	\centering
\tikzset{pill/.style={minimum width=6mm,minimum height=3mm,rounded
corners=1.5mm,draw},
reactor/.style={circle,draw,minimum size=4mm,path picture={
\draw (-2mm,0) -- (2mm,0) (0,-2mm) -- (0,2mm);
\fill (0,0) -- (2mm,0) arc(0:-90:2mm) -- cycle;
\fill (0,0) -- (-2mm,0) arc(180:90:2mm) -- cycle;
}}}
 \begin{tikzpicture}
  \tikzmath{\t1 =160; \w=1;
  \x1 =  2;
  \x2 = 2.3;
  \rr = 4;
    }
    \coordinate (O) at (0,0);
    \draw[fill=lightgray] (O) circle (\rr);
\draw[thick,{Triangle[length=2mm]}-{Triangle[length=2mm]}] (0,\rr) coordinate (Y) -- (O) -- (\rr,0)
coordinate (X);

\draw[thick,dashed] (\x1,\x2) coordinate (F1) --++ (90+\t1:\w) coordinate (F2);
\draw[thick,dashed] (F1) -- ++ (\t1:.4) node[pos=0,sloped,pill,solid]{};
\draw[thick,dashed] (F2) -- ++ (\t1:.4) node[pos=0,sloped,pill,solid]{};
\draw[thick] (F1)  -- (F2) node[pos=0.50,sloped,reactor] (M){~};
\draw[thick,dotted] (O |- M.center) coordinate (y) node[left]{$y$} -- (M.center) coordinate (MM) -- (O -| M.center) coordinate (x) node[below] {$x$}
(M.center) --++(1.5,0) coordinate (BX);
\draw[red,thick,-latex] (M.center) -- ($(M.center)!1.5cm!-90:(F2)$) coordinate (v) node[above]  {$v$};
\draw pic ["$\theta$",draw,-,thick,angle radius=1cm,angle eccentricity=1.2] {angle = BX--MM--v};
\draw[thick] (O) -- node [above left] {$r$} (M.center); 
\draw pic ["$\phi$",draw,-,thick,angle radius=1cm,angle eccentricity=1.2] {angle = x--O--MM};
\node at ($(O) - (0,\rr-1)$) {Escape Region $\Omega$};
\end{tikzpicture}
	\caption{Coordinate system and escape region (shaded).}
	\label{fig:fig1}
\end{figure}

Consider a robot moving with constant speed $v > 0$ and kinematics described by a Dubins car in the two-dimensional Euclidean plane as shown in Fig.~\ref{fig:fig1}.
Specifically, the robot's position $(x,y) \in \mathbb{R} \times \mathbb{R}$ and heading angle $\theta \in (-\pi, \pi]$ evolve according to\footnote{We use dot notation to denote derivatives with respect to time.}
\begin{align}
	\label{eq:cartesianDynamics}
	\begin{aligned}
		\dot{x}(t)
		&= v \cos \theta(t)\\
		\dot{y}(t)
		&= v \sin \theta(t)\\
		\dot{\theta}(t)
		&= - \omega u(t)
	\end{aligned}
\end{align}
for (continuous) time $t \geq 0$ where $\omega > 0$ is a constant maximum turn rate in radians per second, and the robot's control input is the normalized turn rate $u(t)$ taking values in the interval $[-1,1]$.
Here, controls $u(t) \in (0,1]$ correspond to right-hand turns, controls $u(t) \in [-1,0)$ correspond to left-hand turns, and $u(t) = 0$ corresponds to no turning motion.

Suppose that the robot initially starts at some location in a circle with radius $\rho > 0$ and centered, without loss of generality, at the origin.
We shall denote the circular region as $\Omega \triangleq \{ (x,y) \in \mathbb{R} \times \mathbb{R} : x^2 + y^2 < \rho^2 \}$, and refer to it as the \emph{escape region}.
Our minimum-time escape problem is to find a turning strategy that enables the robot to escape from $\Omega$ in minimum time.

To formulate our escape problem mathematically, let us define the state vector $\bx(t) \triangleq [x(t) \; y(t) \; \theta(t)]'$ and write the robot's equations of motion \eqref{eq:cartesianDynamics} in state-space form as
\begin{align}
	\dot{\bx}(t)
	= f \left( \bx(t), u(t) \right)
\end{align}
where
\begin{align*}
	f \left( \bx(t), u(t) \right)
	&\triangleq \begin{bmatrix}
		v \cos \theta(t)\\
		v \sin \theta(t)\\
		- \omega u(t)
	\end{bmatrix}.
\end{align*}
Let us also define $r(t) \geq 0$ as the range of the robot from the center of the escape region at time $t$, i.e., the radial distance of the robot from the origin given by
\begin{align*}
    r(t)
    &\triangleq \sqrt{x^2(t) + y^2(t)}.
\end{align*}
Our minimum-time escape problem is then
\begin{align}
	\label{eq:ocProblem}
	\begin{aligned}
		&\underset{}{{\inf}}  & & \int_0^T 1 \; dt\\
		&\mathrm{s.t.} & & \dot{\bx}(t) = f(\bx(t), u(t)) \\
	    & & & \bx(0) = \bx_0 \in \mathbb{R}^3\\
		& & & r(0) < \rho\\
		& & & u(t) \in [-1,1] \text{ for all } t \in [0,T]\\
		& & & r(T) = \rho\\
		& & & \dot{r}(T) \geq 0
	\end{aligned}
\end{align}
where the optimization is over control functions $u : [0,T] \mapsto [-1,1]$ and the final time $T > 0$.
Here, the initial state constraint $r(0) < \rho$ corresponds to the robot starting at some location in the escape region $\Omega$, and the terminal state constraints $r(T) = \rho$ and $\dot{r}(T) \geq 0$ encode that our concept of escape is when the robot reaches the boundary of the region $\Omega$ and either remains on it, or continues away from it.
Hence, \eqref{eq:ocProblem} corresponds to a minimum-time final-time-free optimal control problem with control constraints and the terminal state constrained to a target set.

In this paper, we seek to solve our minimum-time escape problem \eqref{eq:ocProblem} by identifying a closed-form optimal feedback control law that expresses the optimal controls $u^*(t)$ at each time $t \in [0,T]$ as a function of the current state $\bx(t)$.
We highlight that even in the relatively simple setting of optimally controlling a Dubins car, such closed-form optimal feedback control laws are rare (see e.g., \citep{Chen2020,Shima2020}).
We note, however, that our minimum-time escape problem differs subtly from a standard Dubins car problem in that the terminal state $\bx(T)$ is not fully specified and instead only constrained to satisfy $r(T) = \rho$ and $\dot{r}(T) \geq 0$.

\section{Optimal Control Solution}
\label{sec:ocsol}

In this section, we use Pontryagin's minimum principle for minimum-time final-time-free optimal control problems to identify a simple feedback control law solving \eqref{eq:ocProblem}.

\subsection{Structure of Solution via the Minimum Principle}
To state the conditions of the minimum principle as applicable to \eqref{eq:ocProblem}, let us define the \emph{Hamiltonian} function
\begin{align}\notag
	&H \left( \bx(t), u(t), \blambda(t) \right)\\\notag
	&\triangleq 1 + \blambda'(t) f \left( \bx(t), u(t) \right)\\\label{eq:ham}
	&= 1 
	+ \lambda_x(t) v \cos \theta(t)
	+ \lambda_y(t) v \sin \theta(t)
	- \lambda_\theta(t) \omega u(t)
\end{align}
where $\blambda(t) \triangleq [ \lambda_x(t) \; \lambda_y(t) \; \lambda_\theta(t)]'$ is the costate (or adjoint) vector-valued function with components $\lambda_x : [0,T] \mapsto \mathbb{R}$, $\lambda_y : [0,T] \mapsto \mathbb{R}$, and $\lambda_\theta : [0,T] \mapsto \mathbb{R}$.\footnote{We use $\prime$ to denote the (matrix) transpose operator.}
We then have the following lemma.

\begin{lem}
\label{lemma:mp}
If $u^* : [0,T] \mapsto [-1,1]$ solves \eqref{eq:ocProblem} without the constraint $\dot{r}(T) \geq 0$ then there exists associated state $\bx^*$ and costate $\lambda$ functions satisfying:
\begin{enumerate}[a)]
    \item The Hamiltonian minimum condition that
	\begin{align}
	    \label{eq:minHam}
		H \left( \bx^*(t), u^*(t), \blambda(t) \right)
		&\leq H \left( \bx^*(t), u(t), \blambda(t) \right)
\end{align}
for all $u(t) \in [-1,1]$ and all $t \in [0,T]$;
\item
The costate differential equations
\begin{align}
	\label{eq:costate}
	\begin{split}
		\dot{\lambda}_x(t)
		&= -\pdv{H \left( \bx^*(t), u^*(t), \blambda(t) \right)}{x}
		= 0\\
		\dot{\lambda}_y(t)
		&= -\pdv{H \left( \bx^*(t), u^*(t), \blambda(t) \right)}{y}
		= 0\\
		\dot{\lambda}_\theta(t)
		&= -\pdv{H \left( \bx^*(t), u^*(t), \blambda(t) \right)}{\theta}\\
		&= \lambda_x(t) v \sin \theta(t) - \lambda_y(t) v \cos \theta(t)
	\end{split}
\end{align}
for $t \in [0,T]$; and,
\item
The transversality conditions
\begin{align}
	\label{eq:travCondA}
	H \left( \bx^*(T), u^*(T), \blambda(T) \right)
	= 0
\end{align}
and
\begin{align}
    \label{eq:travCondB}
    \blambda(T)
	= \mu \begin{bmatrix}
			x(T) &
			y(T) &
			0
 \end{bmatrix}'
\end{align}
where $\mu \in \mathbb{R}$ is some real-valued constant.
\end{enumerate}
\end{lem}
\begin{pf}
The proof is a straightforward application of the minimum-principle conditions presented in \citep[p.\ 233]{Kirk2004} (see also \citep[p.\ 215]{Lewis2012a}) to \eqref{eq:ocProblem} and \eqref{eq:ham} without the constraint $\dot{r}(T) \geq 0$.
The transversality conditions \eqref{eq:travCondA} and \eqref{eq:travCondB} in particular hold because the final state $\bx^*(T)$ must belong to the set (or surface) defined by $0.5r^2(T) - 0.5\rho^2 = 0$.
The proof is complete.
\hspace*{\fill} \qed
\end{pf}

In Lemma \ref{lemma:mp}, we omitted explicit consideration of the constraint $\dot{r}(T) \geq 0$ to avoid technicalities dealing with terminal state constraints.
Our approach will therefore be to initially proceed with solving \eqref{eq:ocProblem} without this terminal state constraint, but then later use it to exclude certain candidate solutions.
Following this approach, our next result establishes the form of the optimal controls, costates, and Hamiltonian without the constraint $\dot{r}(T) \geq 0$.

\begin{lem}
	\label{lemma:optimalControlsCostates}
	Let us define the constant $\beta \triangleq v \mu \rho \in \mathbb{R}$, and the polar coordinates $(r(t), \phi(t))$ with range $r(t)$ and azimuth angle $\phi(t) \in (-\pi,\pi]$ such that
\begin{align*}
	x(t) = r(t) \cos \phi(t) \text{ and } y(t) = r(t) \sin \phi(t)
\end{align*}
for $t \in [0,T]$.
	The optimal controls $u^* : [0,T] \mapsto [-1,1] $ solving \eqref{eq:ocProblem} without the constraint $\dot{r}(T) \geq 0$ then satisfy
	\begin{align}\notag
	&H \left( \bx^*(t), u^*(t), \blambda(t) \right)\\\label{eq:hamZero}
	&\quad= 0\\\label{eq:hamAngles}
	&\quad= 1 + \beta \cos \left( \theta(t) - \phi(T) \right) - \lambda_\theta(t) \omega u^*(t)
\end{align}
for all $t \in [0,T]$, and
	\begin{align}
	    \label{eq:switchLaw}
		u^*(t)
		= \sign \left( \lambda_\theta(t) \right)
		\triangleq \begin{cases}
			1 & \lambda_\theta(t) > 0\\
			0 & \lambda_\theta(t) = 0\\
			-1 & \lambda_\theta(t) < 0
		\end{cases}
	\end{align}
	for all $t \in [0,T]$, with $\lambda_\theta : [0,T] \mapsto \mathbb{R}$ being the solution to the differential equation
\begin{align}
	\label{eq:costateAngles}
	\dot{\lambda}_\theta(t)
	&= \beta \sin \left( \theta(t) - \phi(T) \right)
\end{align}
with $\lambda_\theta(T) = 0$.
\end{lem}
\begin{pf}
The Hamiltonian \eqref{eq:ham} is not an explicit function of time so $\dot{H} \left( \bx^*(t), u^*(t), \blambda(t) \right) = 0$ for all $t \in [0,T]$.
The first transversality condition \eqref{eq:travCondA} thus implies \eqref{eq:hamZero} (see also \citep[p. 116]{Lewis2012a}).

Evaluation of the second transversality condition \eqref{eq:travCondB} yields that $\lambda_x(T) = \mu x(T)$, $\lambda_y (T) = \mu y(T)$, and $\lambda_\theta(T) = 0$.
The costates $\lambda_x(t)$ and $\lambda_y(t)$ are constant for all $t \in [0,T]$ via \eqref{eq:costate}, and so $\lambda_x(t) = \mu x(T)$ and $\lambda_y(t) = \mu y(T)$ for all $t \in [0,T]$.
Substituting these values into the expression for $\dot{\lambda}_\theta(t)$ in \eqref{eq:costate} and recalling that $x(T) = \rho \cos \phi(T)$ and $y(T) = \rho \sin \phi(T)$ due to the definition of the azimuth angle $\phi(T)$ and the terminal constraint that $r(T) = \rho$ gives
\begin{align*}
    \dot{\lambda}_\theta(t)
    &= \beta \left[ \cos\phi(T) \sin \theta(t) - \sin\phi(T) \cos\theta(t) \right]
\end{align*}
for all $t \in [0,T]$.
The differential equation for $\lambda_\theta$ given in \eqref{eq:costateAngles} follows via trigonometric identities.
A similar substitution of $\lambda_x$ and $\lambda_y$ into \eqref{eq:ham} establishes \eqref{eq:hamAngles}.

Let us now establish the switching law \eqref{eq:switchLaw}.
	Substituting \eqref{eq:hamAngles} into \eqref{eq:minHam} and canceling common terms we have that
	\begin{align*}
		\lambda_\theta(t) u^*(t)
		&\geq \lambda_\theta(t) u(t)
	\end{align*}
	for all $t \in [0,T]$ and all admissible controls $u(t) \in [-1,1]$.
	The optimal controls $u^*$ therefore satisfy
	\begin{align*}
		u^*(t)
		&= \begin{cases}
			1 & \lambda_\theta(t) > 0\\
			\text{undetermined} & \lambda_\theta(t) = 0\\
			-1 &  \lambda_\theta(t) < 0
		\end{cases}
	\end{align*}
	for all $t \in [0,T]$ where $\lambda_\theta$ acts as a function for switching between controls.
	Specifically, at isolated times when $\lambda_\theta(t) = 0$ but $\dot{\lambda}_\theta(t) \neq 0$, the controls switch between the two extremes $1$ and $-1$.
	The controls are undetermined, and said to be \emph{singular}, on intervals (non-isolated times) $t \in [t_1,t_2] \subset [0,T]$ when $\lambda_\theta(t)$ and all its time derivatives (if they exist) vanish.
	
	To determine the singular controls, suppose that the controls are singular on some interval $[t_1,t_2]$, then the second time-derivative of $\lambda_\theta(t)$ satisfies
	\begin{align*}
		\ddot{\lambda}_\theta(t)
		= \dot{\theta}(t) \beta \cos \left( \theta(t) - \phi(T) \right)
		= 0 
	\end{align*}
	for all $t \in [t_1, t_2]$ and so either $\beta \cos \left( \theta(t) - \phi(T) \right) = 0$ or $\dot{\theta}(t) = 0$ for all $t \in [t_1, t_2]$.
	However, by using \eqref{eq:hamAngles} and recalling that $\lambda_\theta(t) = 0$ on singular intervals $[t_1, t_2]$, $\beta \cos \left( \theta(t) - \phi(T) \right) = 0$ leads to the contradiction $1 = 0$ in \eqref{eq:hamZero} on $[t_1,t_2]$. 
	We therefore conclude that $\dot{\theta}(t) = 0$ must hold for all $t \in [t_1,t_2]$, and it follows from \eqref{eq:cartesianDynamics} that the singular controls must satisfy $u^*(t) = 0$ for all $t \in [t_1,t_2]$.
	The proof is complete. \hspace*{\fill} \qed
\end{pf}

\subsection{Main Result}
Lemma \ref{lemma:optimalControlsCostates} expresses the optimal controls $u^*$ solving \eqref{eq:ocProblem} without consideration of the constraint $\dot{r}(T) \geq 0$ as a function of the costate function $\lambda_\theta$.
In the following theorem, we establish our main result that the optimal controls $u^*$ solving \eqref{eq:ocProblem} and satisfying the constraint $\dot{r}(T) \geq 0$ are given by a simple feedback control law involving the heading and azimuth angles, $\theta(t)$ and $\phi(t)$.

\begin{thm}
    \label{theorem:op}
	The optimal controls solving \eqref{eq:ocProblem} (with the constraint $\dot{r}(T) \geq 0$) satisfy the feedback control law
	\begin{align}
	\label{eq:optimal}
		u^*(t)
		= \sign \left( \theta(t) - \phi(t) \right)
		\triangleq \begin{cases}
			1 & \theta(t) - \phi(t) > 0\\
			0 & \theta(t) - \phi(t) = 0\\
			-1 & \theta(t) - \phi(t) < 0
		\end{cases}
	\end{align}
	for all $t \in [0,T]$ where the differences between angles are wrapped to $(-\pi,\pi]$.
\end{thm}
\begin{pf}
	We first show that the constraint $\dot{r}(T) \geq 0$ implies that $\beta < 0$ for all candidate solutions of interest.
	To do so, note that the definition of $r(T)$ along with $x(T) = r(T) \cos \phi(T)$, $y(T) = r(T) \sin \phi(T)$, and \eqref{eq:cartesianDynamics}, implies that $\dot{r}(T) = v \cos \left( \theta(T) - \phi(T) \right)$.
	Thus, the constraint $\dot{r}(T) \geq 0$ implies that we are only interested in terminal heading and azimuth angles that satisfy $\cos \left( \theta(T) - \phi(T) \right) \geq 0$.
	Furthermore, \eqref{eq:hamZero} and \eqref{eq:hamAngles} combined with $\lambda_\theta(T) = 0$ give
	\begin{align*}
	    \beta \cos \left( \theta(T) - \phi(T) \right) = -1,
	\end{align*}
	and since $\cos \left( \theta(T) - \phi(T) \right) \geq 0$, we have that $\beta < 0$ under the constraint $\dot{r}(T) \geq 0$.

	Turning our attention to proving the second condition in \eqref{eq:optimal}, we note that Lemma \ref{lemma:optimalControlsCostates} and its proof imply that $u^*(t) = 0$ only occurs concurrently with $\lambda_\theta(t) = 0$ and $\dot{\lambda}_\theta(t) = 0$ due to it being a singular control.
	To determine states (rather than costates) at which $u^*(t) = 0$, it thus suffices to find states at which $u^*(t) = 0$, $\lambda_\theta(t) = 0$, and $\dot{\lambda}_\theta(t) = 0$ occur simultaneously.
	Note first that if $\lambda_\theta(t) = 0$, then $\dot{\lambda}_\theta(t) = 0$ occurs when $\theta(t) - \phi(T) = 0$ or $\theta(t) - \phi(T) = \pi$ (wrapping the difference to $(-\pi,\pi]$) since \eqref{eq:costateAngles} gives that $\dot{\lambda}_\theta(t) = 0$ can be written as
	\begin{align}
	\label{eq:dotLambda}
		\dot{\lambda}_\theta(t)
		= \beta \sin \left( \theta(t) - \phi(T) \right)
		= 0
	\end{align}
	and $\beta < 0$ implies that $\sin \left( \theta(t) - \phi(T) \right) = 0$.
	Furthermore, when $u^*(t) = 0$, $\lambda_\theta(t) = 0$, and $\dot{\lambda}_\theta(t) = 0$ occur simultaneously, we must also have that
	\begin{align*}
	    H \left( \bx^*(t), u^*(t), \blambda(t) \right)
	    = H \left( \bx^*(T), u^*(T), \blambda(T) \right) 
	    = 0
	\end{align*}
	 via \eqref{eq:hamZero} and \eqref{eq:hamAngles}, and so
	\begin{align}
		\label{eq:switchCond}
		\cos \left( \theta(t) - \phi(T) \right)
		&= \cos \left( \theta(T) - \phi(T) \right)
	\end{align}
	since $\lambda_\theta(T) = 0$ via Lemma \ref{lemma:optimalControlsCostates}.
	Recalling that $\dot{r}(T) \geq 0$ implies that $\cos \left( \theta(T) - \phi(T) \right) \geq 0$, it follows that $\theta(t) - \phi(T) = 0$ in \eqref{eq:switchCond} (and not $\theta(t) - \phi(T) = \pi$).
	Hence, $u^*(t) = 0$ if $\theta(t) = \phi(T)$.
	
	Suppose now that $\theta(t_1) = \phi(T)$ at some time $t_1 \in [0,T]$.
	Then $u^*(t) = 0$ if $\theta(t) = \phi(T)$ and the singular nature of $u^*(t) = 0$ implies that $u^*(t_1) = 0$ and $\lambda_\theta(t_1)$ along with all its time derivates (if they exist) vanish.
	Furthermore, \eqref{eq:cartesianDynamics} implies that $\dot{\theta}(t_1) = 0$.
	The optimal heading angle $\theta(t)$ will thus not deviate from $\theta(t) = \theta(t_1) = \phi(T)$ for all $t \in [t_1, T]$, and $\theta(T) = \phi(T)$.
	The robot's optimal trajectory for times $t \in [t_1,T]$ therefore corresponds to following the radial line connecting the origin to the terminal position  $(x(T),y(T))$ on the boundary of $\Omega$.
	Noting that the robot is aligned with a radial line whenever $\theta(t) = \phi(t)$, it follows that $u^*(t) = 0$ if $\theta(t) = \phi(t)$ (with $\theta(t) = \phi(T) = \phi(t)$ under $u^*(t) = 0$), proving the second condition in \eqref{eq:optimal}.
	
To prove the first and third conditions in \eqref{eq:optimal}, define $t_0 \triangleq \min \{t_1, T\}$ where $t_1 \triangleq \inf \{t \in [0,T] : \theta(t) = \phi(t)\}$ with $\inf \emptyset \triangleq \infty$.
	The time $t_0$ thus corresponds to either $T$, or the time $t_1$ at which the controls switch to (and remain) $u^*(t) = 0$ for all $t \in [t_0, T]$; in either case, $\lambda_\theta(t_0) = 0$.
	Prior to $t_0$, either $\theta(t) - \phi(t) > 0$ for all $t \in [0,t_0]$ or $\theta(t) - \phi(t) < 0$ for all $t \in [0,t_0]$ since $\theta(t) - \phi(t)$ cannot change sign without $\theta(t) = \phi(t)$ at $t = t_1$ due to $\theta$ and $\phi$ being continuous in $t$ via existence of their derivatives.
	
Considering $\theta(t) - \phi(t) > 0$ for all $t \in [0,t_0]$, then \eqref{eq:costateAngles} and $\beta < 0$ give $\dot{\lambda}_\theta(t) < 0$ for all $t \in [0,t_0]$.
	Since $\lambda_\theta(t_0) = 0$ and $\dot{\lambda}_\theta(t) < 0$ for all $t \in [0,t_0]$, it follows that $\lambda_\theta(t) > 0$ and thus $u^*(t) = 1$ for all $t \in [0,t_0]$ via Lemma \ref{lemma:optimalControlsCostates}, proving the first condition in \eqref{eq:optimal}.
	Similarly, when $\theta(t) - \phi(t) < 0$ for all $t \in [0,t_0]$, then \eqref{eq:costateAngles} and $\beta < 0$ give $\dot{\lambda}_\theta(t) > 0$ for all $t \in [0,t_0]$.
	Since $\lambda_\theta(t_0) = 0$ and $\dot{\lambda}_\theta(t) > 0$ for all $t \in [0,t_0]$, it follows that $\lambda_\theta(t) < 0$ and thus $u^*(t) = -1$ for all $t \in [0,t_0]$ via Lemma \ref{lemma:optimalControlsCostates}, proving the third condition in \eqref{eq:optimal}.
	The proof is complete. \hspace*{\fill} \qed
\end{pf}

The following corollary characterizes the minimum-time paths, and opens discussion of their geometry.

\begin{cor}
	\label{cor:pathSegments}
	The minimum-time path (i.e., the robot position trajectory) solving \eqref{eq:ocProblem} is comprised of either a straight line, an arc, or an arc followed by a straight line.
\end{cor}
\begin{pf}
    The optimal path is a straight line if $\theta(0) = \phi(0)$ since Theorem \ref{theorem:op} and \eqref{eq:cartesianDynamics} imply $u^*(t) = 0$ and $\dot{\theta}(t) = 0$ for all $t \in [0,T]$.
    The optimal path is either an arc, or an arc followed by a straight line, if $\theta(0) - \phi(0) > 0$ or $\theta(0) - \phi(0) < 0$ since Theorem \ref{theorem:op} and \eqref{eq:cartesianDynamics} imply $u^*(t) = \pm 1$ and $\dot{\theta}(t) \lessgtr 0$ until either $t = T$, or $\theta(t) = \phi(t)$ occurs and the controls switch to $u^*(t) = 0$ until $t = T$.
    \hspace*{\fill} \qed
\end{pf}

\section{Geometric Interpretation}
\label{sec:geo}


In this section, we discuss the geometry of the optimal paths solving \eqref{eq:ocProblem}.
Our discussion is based on the observation that as the robot travels with a constant speed, the problem of finding a minimum-time path is identical to the problem of finding a minimum-length path. 
This re-framing of the problem allows us to draw on the existing rich understanding that emanates from the study of minimum-length paths and particularly the Dubins path.

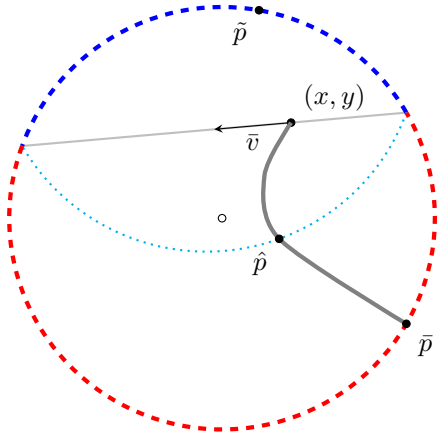
\begin{figure}
    \centering
   \begin{tikzpicture}
[
    scale=.7,
    >=stealth,
    point/.style = {draw, circle,  fill = black, inner sep = 1pt},
    orig/.style   = {draw, circle, inner sep = 1pt},
  ]
\tikzmath{\t1 =30; \t2 =160; \t3 = \t1+360;
\r= 4;
coordinate \x;
\x1 = ({cos(\t1)*\r}, {sin(\t1)*\r});
\x2 = ({cos(\t2)*\r}, {sin(\t2)*\r});
    }

 \draw[lightgray,thick] (\x1) -- (\x2);
 \draw[dashed,blue,ultra thick] (\t1:\r) arc (\t1:\t2:\r);
 \draw[dotted,cyan,thick] (\t1:\r) arc (\t2-180:\t1-180:\r);
 \draw[dashed,red,ultra thick] (\t2:\r) arc (\t2:\t3:\r);
 \draw[->] ($ (\x1) !.3 ! (\x2) $)
    -- node [below] {$\bar{v}$}
    ($ (\x1) ! .5 ! (\x2) $);
    
    \node (robot) at ($ (\x1) !.3 ! (\x2) $) [point, label = {above right:$(x,y)$}]{};
    \coordinate (int) at ($(\t1:\r) + (\t2:\r) + (\t1-180+80:\r)$) ;
     \coordinate (int2) at ($(\t1:\r) + (\t1+180:\r) + (\t2-80:\r)$) ;
     \coordinate (origin) at ($(\t1:\r) + (\t1+180:\r)$) ;
    \draw [gray,ultra thick] plot [smooth] coordinates {(robot)  ($(robot) + (-.5,-1)$)  (int)  (-30:\r) };
	 \node (endpoint) at (-30:\r) [point, label = {below right:$\bar{p}$}]{};
	 
	 \node at (int) [point, label = {below left:$\hat{p}$}]{};
	 \node at (int2) [point, label = {below left:$\tilde{p}$}]{};
	 \node at (origin) [orig]{};

\end{tikzpicture}
    \caption{The scenario described in the proof of Lemma \ref{lem:shorter_arc}.}
    \label{fig:shorter_arc}
\end{figure}

We begin by noting that the chord of the circular escape region $\Omega$ whose normal vector is orthogonal to the vector $\bar{v} \triangleq [v\cos\theta(0) \; v\sin\theta(0)]'$ partitions the boundary of $\Omega$ into two arcs (illustrated in Fig.~\ref{fig:shorter_arc} with $x(0) = x$ and $y(0) = y$).
The following lemma establishes the relationship between the minimum-time (or minimum-length) escape path solving \eqref{eq:ocProblem} and these two arcs.

\begin{lem}\label{lem:shorter_arc}
The end point of the minimum-time escape path lies on the arc of shorter length.
\end{lem}
\begin{pf}
We provide a simple geometric argument using Fig.~\ref{fig:shorter_arc}. To obtain a contradiction assume that the end point of the minimum-time escape path is on the longer arc (dashed red arc in Fig.~\ref{fig:shorter_arc}). Denote this point by $\bar{p}$. By construction, the path connecting $(x,y)$ to $\bar{p}$ intersects an arc (dotted cyan arc in Fig.~\ref{fig:shorter_arc}) obtained from mirroring the shorter arc (dashed blue arc in Fig.~\ref{fig:shorter_arc}) with respect to the chord whose normal vector is orthogonal to $\bar{v}$. Call this intersection point $\hat{p}$. Obviously, the length of the path starting from $(x,y)$ and ending in $\hat{p}$ is shorter than the one ending in $\bar{p}$. Mirroring $\hat{p}$ with respect to the aforementioned chord we obtain another point on the circle, labeled $\tilde{p}$. This leads to a contradiction as we obtain an exit point $\tilde{p}$ on the shorter arc that leads to a shorter path than the exit point $\bar{p}$ on the longer arc. 
\hspace*{\fill} \qed
\end{pf}

For the simplicity of presentation, we denote the shorter arc (blue-dashed line in Fig.~\ref{fig:shorter_arc}) by $\mathcal{A}$ and make the following assumption\footnote{While this assumption might bring about a picture of Derek Zoolander and his inability to turn left, it is by no means restrictive. Without this assumption one needs to introduce two cases for the shorter arc to be on the right and the left and that will lead to a proliferation of indices. The Zoolander reference was included to the chagrin of the first author.}.
\begin{assumption} \label{asmp:RHT}
For all $p\in\mathcal{A}$, $\langle \bar{v}^\perp ,p\rangle \geq 0$ where $\langle \cdot ,\cdot \rangle$ denotes the inner product (with $p$ treated as a vector) and $\bar{v}^\perp$ is obtained by a clockwise rotation of $\bar{v}$ by $\pi/2$.
\end{assumption}

To construct the family of shortest (minimum time or distance) escape paths, we rely on Corollary \ref{cor:pathSegments}. In geometric terms, the corollary characterizes the shortest curve that connects two points in the two-dimensional Euclidean plane with a constraint on the curvature of the path, with only prescribed initial tangent to the path, and unconstrained terminal tangent.

\begin{remark}
It is envisaged that Corollary \ref{cor:pathSegments} can be proven through investigating the homotopy classes of bounded curvature paths via an approach similar to that of \cite{ayala_geometric_2018}. However, that analysis is left as a future work.
\end{remark}

    

Consider now a circle of radius $\varrho$ centered at $c$, denoted by $\mathcal{C}$. From a point $p$, two lines are incident tangent to $\mathcal{C}$ and these tangents touch the circle at points $\tau_1(p)$ and $\tau_2(p)$ given by
$$
\tau_i(p) = c + \dfrac{\varrho^2}{\|p - c\|^2} (p-c)  \pm \dfrac{\rho}{\|p-c\|^2} \left ( \sqrt{\|p-c\|^2 - \varrho^2 } \right ) q,
$$
where $i\in\{1,2\}$ and $q$ is orthogonal to and has the same length as $p - c$.

In our problem, the circle corresponds to the circle of smallest turn radius for the robot given its position and heading. Consequently, $$c \triangleq [ x(0) \; y(0)]' +\varrho \bar{v}^\perp / \|\bar{v}^\perp\|
$$
and $\varrho = v/\omega$. 

\begin{cor}\label{cor:CA0}
Suppose $\mathcal{C}\cap \mathcal{A} = \emptyset$. Under Assumption~\ref{asmp:RHT} the shortest escape path belongs to the collection of all paths obtained by concatenating the arc belonging to $\mathcal{C}$ that starts from $[x(0) \; y(0)]'$ and terminates at $\tau(p)$ and the line segment connecting $\tau(p)$ to $p\in\mathcal{A}$ where $\tau(p) \triangleq \tau_{i^*} (p)$ and $i^* = \arg\min \{\varphi_1(p),\varphi_2(p)\}$ with $\varphi_1(p)$ and $\varphi_2(p)$ being the clockwise angle between $[x(0) \; y(0)]'$ and $\tau_1(p)$ and $\tau_2(p)$.
\end{cor}

Now, we can cast finding the minimum-length escape path as the following optimization problem for the case where $\mathcal{C}\cap \mathcal{A} = \emptyset$:
\begin{align}
    \label{eq:geoOpt}
    \min_{p\in\mathcal{A}} \quad \|p-\tau(p)\| + \varphi(p) \rho,
\end{align}
where $\varphi(p) = \varphi_{i^*} (p)$. 
\begin{rem}
It is worth noting that, $\tau(p)$ corresponds to the $(x,y)$-position where the optimal feedback control law \eqref{eq:optimal} switches to $u^*(t) = 0$.
\end{rem}
We have the following result for the case where $\mathcal{C}$ and $\mathcal{A}$ intersect.
\begin{cor}\label{cor:CA1}
Suppose $\mathcal{C}\cap \mathcal{A} \neq \emptyset$. Under Assumption~\ref{asmp:RHT}, the shortest escape path is obtained by traversing $\mathcal{C}$.
\end{cor}

\begin{rem}
Corollaries \ref{cor:CA0} and \ref{cor:CA1} follow from Corollary \ref{cor:pathSegments} and Lemma \ref{lem:shorter_arc}, and hence, their proofs are omitted.
\end{rem}

\section{Illustrative Simulations}
\label{sec:simulations}
In this section, we present simulations that illustrate the feedback control law \eqref{eq:optimal} and its geometric interpretation.

\begin{figure}
    \centering
    \includegraphics[width = 0.8\columnwidth]{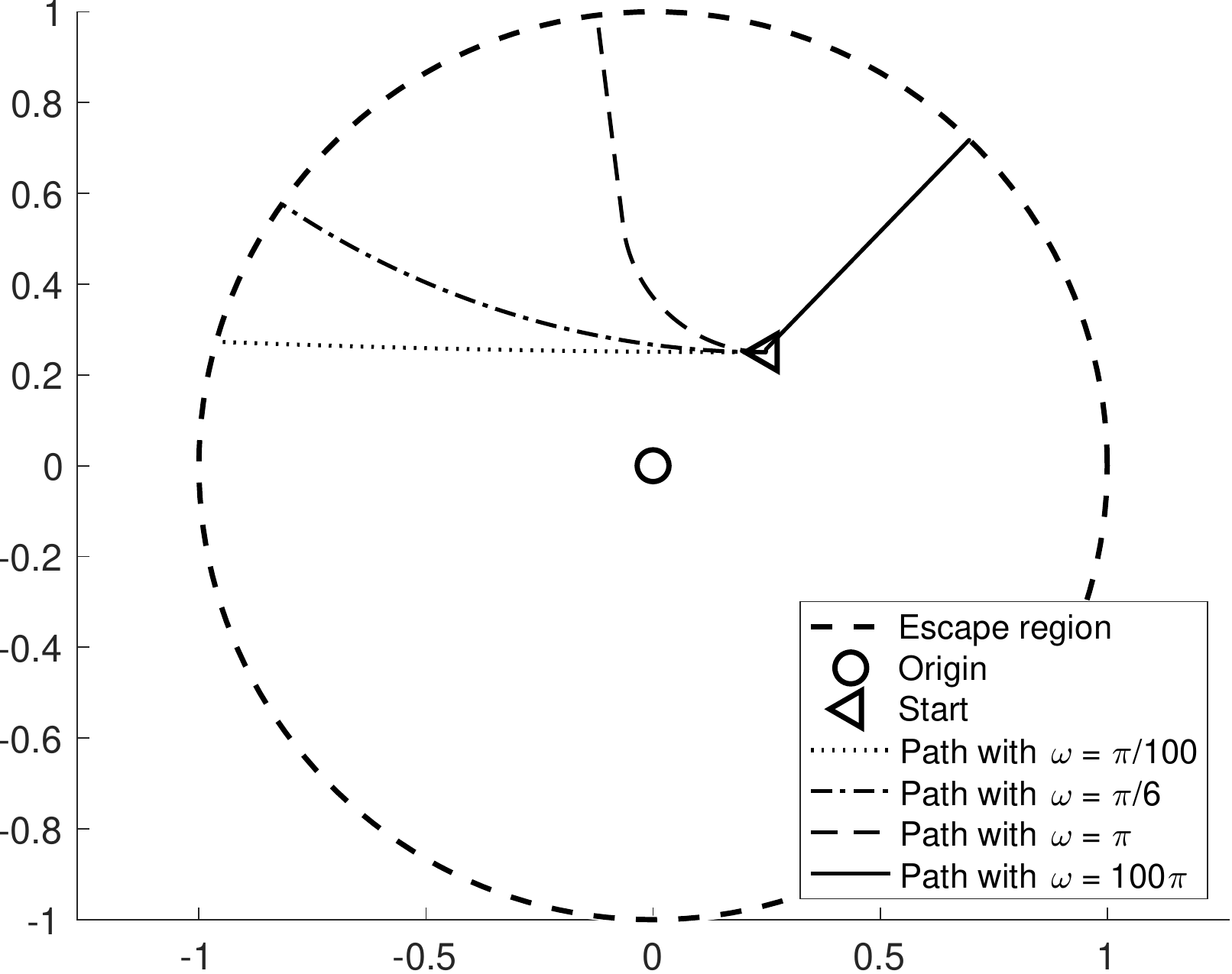}
    \caption{Simulated paths from  $\bx_0 = [0.25 \; 0.25 \; \pi]'$.}
    \label{fig:sim1}
\end{figure}

\begin{figure}
    \centering
    \includegraphics[width = 0.8\columnwidth]{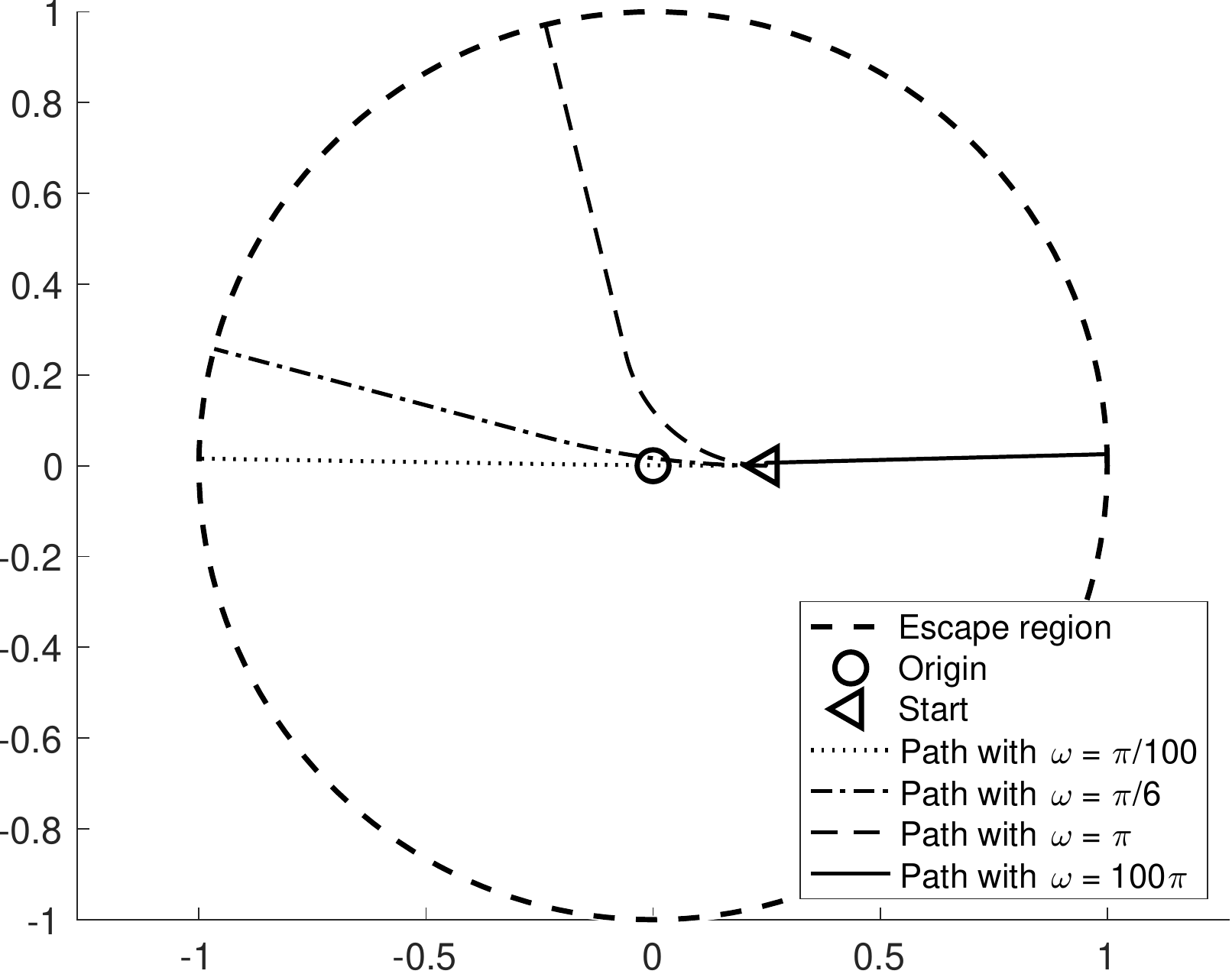}
    \caption{Simulated paths from  $\bx_0 = [0.25\; 0 \; \pi]'$.}
    \label{fig:sim2}
\end{figure}

\begin{figure}
    \centering
    \includegraphics[width = 0.8\columnwidth]{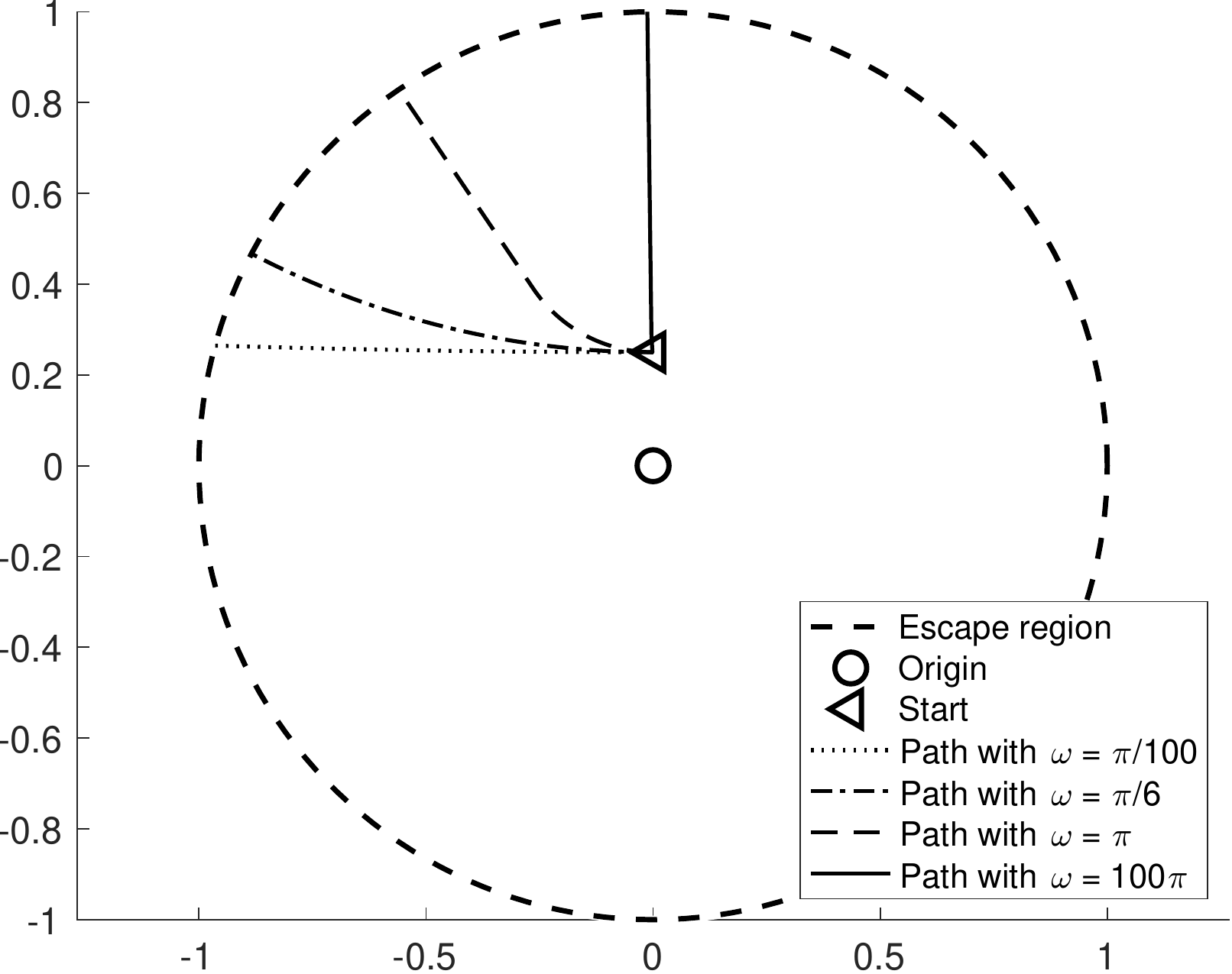}
    \caption{Simulated paths from  $\bx_0 = [0\; 0.5\; \pi]'$.}
    \label{fig:sim3}
\end{figure}

We simulated a robot with $v = 1$m/s attempting to escape from a circle with radius $\rho = 1$m by solving \eqref{eq:cartesianDynamics} with the controls $u$ given by \eqref{eq:optimal} using MATLAB's \texttt{ode45} solver.
We simulated the robot with a variety of maximum turn rates from the set $\omega \in \{\pi/100, \pi/6, \pi, 100\pi\}$ rad/s, and with different initial conditions.
The resulting robot paths are shown in Fig.~\ref{fig:sim1} for $\bx_0 = [0.25\; 0.25\; \pi]'$, Fig.~\ref{fig:sim2} for $\bx_0 = [0.25\; 0\; \pi]'$, and Fig.~\ref{fig:sim3} for $\bx_0 = [0\; 0.25\; \pi]'$.

Figs.~\ref{fig:sim1}--\ref{fig:sim3} illustrate that for small maximum turn rates (large turn radii), the robot's minimum-time path corresponds to simply turning towards the nearest edge of the circle.
In the extreme case of $\omega = \pi/100$ rad/s, this path follows a chord of the escape region.
In contrast, for large maximum turn rates (small turn radii), the robot's minimum-time path corresponds to turning towards the nearest edge of the circle and then switching to tracking a radial line of the escape region.
At the other extreme of $\omega = 100\pi$ rad/s, this path corresponds to almost instantaneously tracking a radial line of the escape region. 

\section{Conclusions and Future Work}
\label{sec:conclusion}

We have posed and solved a minimum-time escape problem for a Dubins car from a circle.
Future work will consider escape problems from non-circular regions and games involving escape regions moving to prevent escape.


\bibliography{Library}


\end{document}